\documentclass[12pt,a4paper]{article}

\usepackage{amssymb,amsmath,color,enumitem,fancyvrb,hyperref,xspace}
\usepackage[affil-it]{authblk}

\fvset{fontsize=\small,commandchars=\\\{\}}

\definecolor{fuchsia}{RGB}{236,0,140}
\definecolor{prompt}{RGB}{245,124,0}
\definecolor{command}{RGB}{48,63,159}

\hypersetup{
    colorlinks,
    urlcolor={fuchsia},
    citecolor={blue}
}

\newcommand{\eps}{\epsilon}
\newcommand{\fuchsia}{\textcolor{fuchsia}{\texttt{Fuchsia}}\xspace}
\newcommand{\linux}{\texttt{Linux}\xspace}
\newcommand{\maple}{\texttt{Maple}\xspace}
\newcommand{\maxima}{\texttt{Maxima}\xspace}
\newcommand{\maximasage}{\texttt{Maxima/Sage}\xspace}
\newcommand{\python}{\texttt{Python}\xspace}
\newcommand{\rank}{\mathrm{rank}}
\newcommand{\sage}{\texttt{SageMath}\xspace}
\newcommand{\code}[1]{\texttt{#1}}
\newcommand{\F}[1]{\texttt{#1}} % use this to style function names
\newcommand{\M}[1]{\mathbb{#1}} % use this to style matrix names
\newcommand{\V}[1]{\mathbf{#1}} % use this to style vector names
\newcommand{\prompt}[2]{\textcolor{prompt}{#1} \textcolor{command}{#2}}
\newcommand{\functionitem}[2]{\item[$\F{#1}(#2)$\hfill\textit{(function)}]}
\newcommand{\classitem}[1]{\item[$#1$\hfill\textit{(class)}]}

  \newlength{\dinwidth} \newlength{\dinmargin}
\title{
  \begin{flushright}
  \tt\normalsize{DESY-16-219}\\ 
  \end{flushright}
  \vspace{1cm}
  \Large \bf \fuchsia: a tool for reducing differential equations for Feynman master integrals to epsilon form\\
  \vspace{0.5cm}
  \normalsize \bf Version 16.11.14
}

\author[a]{Oleksandr Gituliar%
    \thanks{Corresponding author; email address:
        \href{mailto:oleksandr.gituliar@desy.de}
            {oleksandr.gituliar@desy.de}}}

\author[b]{Vitaly Magerya}

\affil[a]{II. Institut f\"ur Theoretische Physik, Universit\"at Hamburg,
Luruper Chaussee 149, D-22761 Hamburg, Germany}
\affil[b]{Novomoskovsk, Ukarine}

\begin{document}

\maketitle
\thispagestyle{empty}

\begin{abstract}
We present \fuchsia\xspace --- an implementation of the Lee algorithm, which for a given system of ordinary differential equations with rational coefficients $\partial_x\,\V f(x,\eps) = \M A(x,\eps)\,\V f(x,\eps)$ finds a basis transformation $\M T(x,\eps)$, i.e., $\V f(x,\eps) = \M T(x,\eps)\,\V g(x,\eps)$, such that the system turns into the \textit{epsilon form}: $\partial_x\, \V g(x,\eps) = \eps\,\M S(x)\,\V g(x,\eps)$, where $\M S(x)$ is a Fuchsian matrix.
A system of this form can be trivially solved in terms of polylogarithms as a Laurent series in the dimensional regulator $\eps$.
That makes the construction of the transformation $\M T(x,\eps)$ crucial for obtaining solutions of the initial system.

In principle, \fuchsia can deal with any regular systems, however its primary task is to reduce differential equations for Feynman master integrals.
It ensures that solutions contain only regular singularities due to the properties of Feynman integrals.
\end{abstract}
\newpage

{\bf PROGRAM SUMMARY}

\begin{small}
\noindent

{\em Program Title:}
    \fuchsia

{\em Authors:}
    O.~Gituliar and V.~Magerya

{\em Program obtainable from:}
    \url{https://github.com/gituliar/fuchsia/}

{\em Journal Reference:}
    %Leave blank, supplied by Elsevier.

{\em Catalog identifier:}
    %Leave blank, supplied by Elsevier.

{\em Licensing provisions:}
    ISC license

{\em Programming language:}
    \python 2.7

{\em Operating system:}
    \linux, \texttt{Unix}-like

{\em RAM:}
    Dependent upon the input data. Expect hundreds of megabytes.


{\em Classification:}
    5 Computer Algebra, 11.1 High Energy Physics and Computing

{\em External routines/libraries:}
    \href{http://www.sagemath.org/}{\sage} (7.0 or higher), \maple (optional)

{\em Nature of problem:}
    Feynman master integrals may be calculated from solutions of a linear system of differential equations with rational coefficients.
    Such a system can be easily solved as an $\eps$-series when its epsilon form is known.
    Hence, a tool which is able to find the epsilon form transformations can be used to evaluate Feynman master integrals. 

{\em Solution method:}
    Methods of Moser \cite{Mos59} and Lee \cite{Lee15}.

{\em Restrictions:}
    Systems of single-variable differential equations are considered, however unknown functions may also depend on other symbolic arguments.
    A system needs to be reducible to Fuchsian form and eigenvalues of its residues must be of the form $n + m\,\eps$, where $n$ is integer.

{\em Running time:}
    Depends upon the input data, its size and complexity.
    Around an hour in total for an example $74\times74$ matrix with 20 singular points on a PC with a 1.7GHz Intel Core i5 CPU.

\end{small}

\newpage

\tableofcontents

\section{Introduction}

More than 60 years have passed since Richard Feynman proposed a diagrammatic approach for calculating perturbative processes in quantum field theories.
Since then Feynman integrals calculus has grown to a separate branch of mathematical physics with a big community of scientists conducting research in this exciting field.
With no doubt we can say that none of the recent discoveries in the high-energy particle physics could happen without precise theoretical calculations, which are based on the Feynman integrals calculation techniques. 
It is also clear that such techniques will play a key role for discoveries at the present and future high-energy colliders, hence their development and further improvement are a very important task.

Recent progress in computational techniques has made it possible to automate the calculation of Feynman integrals; problems which seemed impossible 10 years ago now are successfully solved with state-of-the-art computer algorithms.
Among the most popular are integration-by-parts (IBP) reduction~\cite{CT81} and the method of differential equations (DE)~\cite{Kot91a,Kot91b,Kot91c}; for a detailed overview of these and other methods see~\cite{Smi06}.

In this paper we discuss the method of differential equations.
In particular, we focus on the fact that a solution to the system of DEs may be easily found as an $\eps$-series when an epsilon form of this system is known~\cite{Henn13}.
We consider a general algorithm to find an epsilon form of a given system of differential equations in one variable developed by Lee~\cite{Lee15}, for the review of the subject see \cite{Henn14} and \cite{Pap14,Tan15,ABB15}.
The Lee algorithm allows
\begin{enumerate}
  \item To find a Fuchsian form of the system using improved Moser reduction algorithm \cite{Mos59}; and then
  \item To normalize eigenvalues of the Fuchsian system in all singular points.
\end{enumerate}
If these two steps are successfully completed%
\footnote{In principle, the first step can always be done because Feynman integrals contain only logarithmic singularities. For the discussion of potential complications in the second step see Section~\ref{sec:2}.}
then a transformation which puts the initial system into the epsilon form may be easily found.
Although this method is focused on systems for Feynman integrals, it also may be successfully used for different problems provided that the requirements on particular properties of the initial system are satisfied.
For another method for finding the epsilon form of a given system with multiple scales see \cite{Mey16a,Mey16b}.

Until recently, no implementation of the Lee method was made publicly available.
This fact motivated us to develop \fuchsia \ --- the first public implementation of the Lee algorithm~\cite{Lee15} which was presented in \cite{GM16}.
Another implementation of this method, called \texttt{Epsilon}, was recently presented in \cite{Pra17}.
In this paper we provide a description of some implementation and usage details of \fuchsia which together with algorithms and tools for the integration-by-parts reduction~\cite{Lap00,Smi08,MS12,Lee12,Lee13,SS13,Smi14,GLZ16} form a powerful tandem for evaluating Feynman integrals.

This paper is organized as follows: in Section \ref{sec:2} we introduce notation and definitions followed by a brief review of the Lee method and related algorithms implemented in \fuchsia.
In Section~\ref{sec:3} we describe how to install and how to use \fuchsia from different environments, depending on your goal and programming experience.

\section{Overview of the Lee method}
\label{sec:2}

\subsection{Notation and definitions}

Let us consider a system of ordinary differential equations (ODEs) of this form:
\begin{equation}
\label{eq:dj}
    \partial_x \V J(x,\eps) = \M A(x,\eps)\,\V J,
\end{equation}
where $\V J(x,\eps)$ is a column-vector of $n$ unknown functions (e.g., master integrals);
$x$ is a free variable;
$\eps$ is an infinitesimally small parameter (e.g., a dimensional regulator in $d=4-2\eps$ dimensions);
$\M A(x,\eps)$ is an $n \times n$ matrix, rational in both $x$ and $\eps$.

In the general case $\M A(x,\eps)$ may have a finite number of poles in $x$ at $x\in\{x_k\}$, including a pole at infinity.
The asymptotic behavior of $\M A(x,\eps)$ around these poles can be described as:
\begin{equation}
\label{eq:axeps}
    \M A(x \to x_k,\eps) =
    \begin{cases}
        (\M A_{k0}(\eps) + \M A_{k1}(\eps)\,(x - x_k) + \dots)/(x - x_k)^{1+p_k} & \text{if $x_k\neq\infty$}, \\
        -(\M A_{k0}(\eps) + \M A_{k1}(\eps)\,x^{-1} + \dots)\,x^{-1 + p_k} & \text{if $x_k=\infty$},
    \end{cases}
\end{equation}

where $p_k$ is the \textit{Poincar\'e rank} of $\M A(x,\eps)$ at the singular point $x=x_k$.
If $p_k=0$, we call $\M A(x,\eps)$ \textit{Fuchsian in $x=x_k$}, and $\M A_{k0}$ \textit{matrix residue} of $\M A(x,\eps)$ at $x=x_k$.
If all $p_k=0$, we call $\M A(x,\eps)$ \textit{Fuchsian}.

The behavior of the system at $x=\infty$ is a bit of a special case, but it is essential in the overall reduction process.
One must always keep in mind the $x=\infty$ point, and treat it on the same footing as other singular points.

\paragraph{Equivalent systems.}
We are interested in transforming system~\eqref{eq:dj} into a simpler form.
For this purpose let us consider a change of basis from $\V J$ into $\V J'$ using the linear transformation $\M T(x,\eps)$:
\begin{equation}
  \V J = \M T(x,\eps)\, \V J'.
\end{equation}

This leads to the {\em equivalent system} of ODEs
\begin{equation}
  \partial_x \V J' = \M A'(x,\eps)\,\V J',
\end{equation}
with the new matrix being
\begin{equation}
\label{eq:ta}
  \M A'(x,\eps) = \M T^{-1} \left( \M A \M T - \partial_x \M T \right).
\end{equation}

Generally speaking, the transformations $\M T(x,\eps)$ may have an arbitrary form.
However in the scope of this paper we will require the transformation matrices to be rational in both $x$ and $\eps$.
This restriction guarantees that the equivalent matrix $\M A'(x,\eps)$ and hence all equivalent systems are in rational form, thus making the expansion~\eqref{eq:axeps} possible.

In particular, we will be using the transformation constructed by stepwise application of a \textit{$\M P$-balance between $x=x_1$ and $x=x_2$}, defined as:
\begin{equation}
\label{eq:bal}
    \mathcal{B}(\M P(\eps), x_1, x_2; x) = \M I - \M P(\eps) + c \frac{x-x_1}{x-x_2} \M P(\eps),
\end{equation}
\begin{equation*}
    c \equiv
    \begin{cases}
        1/x_1 & \text{if $x_1=\infty$}, \\
        x_2 & \text{if $x_2=\infty$}, \\
        1 & \text{otherwise},
    \end{cases}
\end{equation*}
where $\M P(\eps)$ is a projector matrix (that is, $\M P^2=\M P$).

\paragraph{Classification of singularities.}
Following \cite{Mos59}, for a system \eqref{eq:dj} and it's Laurent expansion~\eqref{eq:axeps} we define a rational number
\begin{equation}
\label{eq:mk}
    m_k(\M A) = p_k + \frac{rank(\M A_{k0})}{n}
\end{equation}
as the {\em Moser order} of $\M A(x,\eps)$ at point $x=x_k$.

Equivalent systems do not necessarily have identical Moser orders.
In fact \cite{Mos59} introduces an algorithm that constructs a transformation decreasing $rank(\M A_{k0})$ by at least one (thus reducing $m_k(\M A)$),\footnote{Note, that if $rank(\M A_{k0})$ reaches zero, this means that $\M A_{k0}$ is now zero itself, and thus the Poincar\'e rank $p_k$ was decreased by at least one.} or certifies that no further order reduction can be achieved.

Let us then denote the {\em minimal order} of $\M A(x,\eps)$ at $x=x_k$ as:
\begin{equation}
\label{eq:muk}
  \mu_k(\M A) = \min m_k(\M A'), \text{ for } \forall \; \M T.
\end{equation}

If $\mu_k(\M A) < m_k(\M A)$ we say that the matrix $\M A(x,\eps)$ is {\em Moser-reducible} at $x=x_k$.

With this in mind we can classify a point $x=x_k$ of $\M A(x,\eps)$ as:
\begin{itemize}
  \item {\em regular point}, if $m_k(\M A) = 0$;
  \item {\em apparent singularity}, if $m_k(\M A) > 0$ and $\mu_k(\M A) = 0$;
  \item {\em regular singularity}, if $0 < \mu_k(\M A) \le 1$;
  \item {\em irregular singularity}, if $\mu_k(\M A) > 1$.
\end{itemize}
The matrix $\M A(x,\eps)$ is called {\em Fuchsian} if it does not contain irregular singularities at any value of $x$ including $\infty$.

\subsection{Reduction to epsilon form}

In the previous section we have introduced the notation and key definitions related to the Fuchsian theory of ODEs.
Now we are ready to review the reduction method proposed by Lee in~\cite{Lee15}.
With its help we can construct a rational transformation $\M T(x,\eps)$ which converts a system of ordinary differential equations with rational coefficients given by the matrix $\M A(x,\eps)$ to an equivalent system given by the matrix $\M M(x,\eps)$ which is Fuchsian and has an \textit{epsilon form} (also called {\em canonical} in \cite{Henn13}), i.e., $\M M(x,\eps) = \eps\, \M S(x)$.
When the epsilon form $\M M(x,\eps)$ of the initial system $\M A(x,\eps)$ is found we can easily solve it as a Laurent series in $\eps$ and restore solutions for the initial system $\M A(x,\eps)$ --- which is our ultimate goal --- by solving a linear system of equations.

The whole method is performed in the following three steps:

\begin{enumerate}
    \item Given a matrix $\M A(x,\eps)$, find an equivalent system $\M A'(x,\eps)$ and a corresponding transformation $\M T(x,\eps)$, such that $\M A'(x,\eps)$ is Fuchsian. We call this step \textit{fuchsification}.
    \item Given a Fuchsian matrix $\M A(x,\eps)$ with eigenvalues of all its residues of the form $n+m\,\eps$, where $n$ is integer, find an equivalent system (along with the transformation) which is still Fuchsian, but with residue eigenvalues being all of the form $k\,\eps$. We call this step \textit{normalization}.
    \item Given a normalized matrix $\M A(x,\eps)$, find an equivalent matrix $\M A'(x)$ in epsilon form, i.e. such that $\M A'(x,\eps) = \eps \, \M S'(x)$. We call this step \textit{factorization}.
\end{enumerate}

\subsubsection{Fuchsification}
\label{sec:fuchs}

To \textit{fuchsify} a system~\eqref{eq:dj} means to find an equivalent Fuchsian system.
This, of course, is only possible if $\M A(x,\eps)$ has no irregular singularities, or in other words $\mu_k(\M A)\le1$ for all $k$.\footnote{
    We expect this to be often the case in practice, in particular for ODEs corresponding to Feynman integrals which are known to have only logarithmic singularities, hence be solutions of some Fuchsian ODEs.
}

In the case of a single ODE of order~$n$, the minimal Moser order can be computed explicitly from power counting analysis of its coefficients (see the generalization of Fuchs' theorem in~\cite{Mos59}).
This is not possible for ODE systems like~\eqref{eq:dj}.
Instead, we have a criterion for Moser-reducibility of the form:

\textbf{Theorem 1.}
{\em If $m_k(\M A) > 1$ then the system \eqref{eq:dj} is Moser-reducible at $x=x_k$ if and only if the polynomial}
\begin{equation}
\label{eq:red_cond}
    \Delta^{r_k} \det\left(\frac{\M A_{k0}}{\Delta} + \M A_{k1} - \lambda \,\M I\right),
\end{equation}
{\em vanishes identically in $\lambda$ at $x=x_k$, where $r_k=\rank(\M A_{k0})$ and $\Delta=x-x_k$ if $x_k\ne\infty$, or $\Delta=1/x$ if $x_k=\infty$.}

When this condition fails the singularity in $x=x_k$ is {\em irregular}.

In addition to this criterion we have a method for constructing a transformation that lowers the Moser order at $x_i$, provided that the system is Moser-reducible at that point (possibly at the expense of increasing the Poincar\'e rank $p_j$ at another point $x=x_j$ by one).
This is done by selecting a projector matrix $\M P$ equal to a sum of products of a particular subset of (generalized) eigenvectors of $\M A_{i0}$ and $\M A_{j0}$, and constructing a $\M P$-balance between $x_i$ and~$x_j$.

The reader can find the details of this construction in \cite{Lee15}, but it is important to note that even if the system is Moser-reducible at $x_i$, it is only sometimes possible to construct a transformation that lowers $m_i(\M A)$ without increasing Poincar\'e rank at $x_j$.
Sometimes the best we can do is to choose $x_j$ to be some regular point (with $p_j=-1$), and use a transformation that decreases $m_i(\M A)$ at the expense at increasing $p_j$ to $0$, effectively introducing an apparent singularity where there was none before.\footnote{
    In practice these apparent singularities are not a major problem, since they are subsequently removed during the normalization step.
    Still, we try not to introduce them if possible in order to decrease the intermediate expression sizes and to increase the overall performance.
}

With this in mind, to reduce a system to Fuchsian form we need to combine the reducibility check with the Moser rank-lowering transformation in stepwise fashion, as follows:

\begin{enumerate}
    \item Select some point $x_k$ with $m_k(\M A) > 1$. If none exist, reduction is complete.
    \item Check if $\M A(x,\eps)$ is reducible at $x=x_k$. If not, fail.
    \item Find a transformation $\M T(x,\eps) = \mathcal{B}(\M P(\eps), x_k, x_j; x)$ that lowers $m_k(\M A)$.
    \item Apply $\M T(x,\eps)$ and repeat from Step 1.
\end{enumerate}

In \fuchsia, this process is implemented by function \code{fuchsify}; see Section~\ref{sec:usage_py} for it's usage.

Finally, let us mention that a similar problem of reducing Poincar\'e ranks and Moser orders of rational matrices was actively studied by Barkatou and co-authors, e.g., see~\cite{BP99}.
They developed algorithms implementing their method~\cite{BP99} which is available in the standard \texttt{Maple} package \texttt{DEtools} as \code{moser\_reduce} and \code{super\_reduce} routines.

\subsubsection{Normalization}
\label{sec:norm}

To \textit{normalize} a Fuchsian system \eqref{eq:dj} with matrix residue eigenvalues of the form $n+m\,\eps$ (where $n$ is integer), implies to find an equivalent Fuchsian system with residue eigenvalues of the form $m\,\eps$.

Just like in the previous step, the normalizing transformations are found by stepwise application of balance transformations~\eqref{eq:bal}.
We refer the reader to~\cite[p.~11]{Lee15} for the description of how such balances are constructed, but we will note that this transformation is possible due to these two facts:
\begin{itemize}
    \item Given a properly selected projector matrix $\M P(\eps)$, the balance $\mathcal{B}(\M P(\eps), x_i, x_j; x)$ shifts one of the eigenvalues of $\M A_{i0}$ by $\pm1$, shifts one of the eigenvalues of $\M A_{j0}$ by $\mp1$, and does not change the Poincar\'e ranks at any point.
    \item Since the system is Fuchsian, the sum of all its matrix residues is zero, and thus, the sum of all residue eigenvalues is zero as well.
\end{itemize}
Combining these two facts, we perform the normalization by shifting the residue eigenvalues by $1$ at each step, until a state is reached where the integer parts of all the eigenvalues are zero.

In \fuchsia, the normalization step is implemented by the function \code{normalize}.
See Section~\ref{sec:usage_py} for its usage.

It may happen that after fuchsification we obtain a system for which the residue eigenvalues do not fit to the $n+m\,\eps$ form neatly.
In this case it is sometimes possible to rectify the problem by using some non-linear change of variables.
Unfortunately we do not have an automated solution for such cases, and users are expected to find transformations appropriate for their system manually.

\subsubsection{Factorization}
\label{sec:fact}
After the normalization we have obtained a matrix $\M A(x,\eps)$ with all residue eigenvalues of the form $m\,\eps$.
The final step is to \textit{factorize} it by finding an equivalent matrix which is by itself proportional to $\eps$, so $\M A'(x,\eps)=\eps\,\M S(x)$.

A transformation which is constant in $x$ is sufficient for this task.
Let $\M T(\eps)$ be such a transformation, then according to \eqref{eq:ta}, we have:
\begin{equation}
  \M A'(x,\eps) \equiv \eps\, \sum_i \frac{\M S_i}{x-x_i} = \sum_i \M T^{-1}(\eps) \frac{\M A_{i0}(\eps)}{x-x_i} \M T(\eps),
\end{equation}
or explicitly
\begin{equation}
\label{eq:seps}
  \M S_i = \M T^{-1}(\eps) \frac{\M A_{i0}(\eps)}{\eps} \M T(\eps).
\end{equation}
Since $\M S_i$ in \eqref{eq:seps} is the same no matter what values $\eps$ takes, we can say that:
\begin{equation}
  \M S_i = \M T^{-1}(\eps) \frac{\M A_{i0}(\eps)}{\eps} \M T(\eps) =
    \M T^{-1}(\mu) \frac{\M A_{i0}(\mu)}{\eps} \M T(\mu),
\end{equation}
from where we obtain a system of linear equations for $\M T(\eps,\mu) \equiv \M T(\eps) \, \M T^{-1}(\mu)$:
\begin{equation}
  \frac{\M A_{i0}(\eps)}{\eps} \M T(\eps,\mu) = \M T(\eps,\mu) \frac{\M A_{i0}(\mu)}{\mu}, \text{for all $i$.}
\end{equation}
Solving this system for $\M T(\eps,\mu)$ we can reconstruct the initial transformation as $\M T(\eps) = \M T(\eps,\mu_0)$, where $\mu_0$ can be chosen arbitrary as long as $\M T(\eps)$ will come out invertible.

In general, the solution for $\M T(\eps,\mu)$ can have multiple free variables aside from just $\mu$.
We choose to set all of them to random small integers, preferably zeros, which keeps the resulting matrix $\M S(\eps)$ simple.

In \fuchsia the factorization step is implemented as the \code{factorize} routine (again, see Section~\ref{sec:usage_py} for its usage).

\subsubsection{Block-triangular form}
\label{sec:blockreduce}

It often happens that a matrix which defines an ODEs is sparse, i.e. it has many zeros, and can be shuffled into block-triangular form with small blocks.
It is possible to exploit this fact to considerably speed up fuchsification and normalization.

Let us consider a block-triangular matrix of this form:
\begin{equation}
\label{eq:bdiag}
\M A(x,\eps)=
\left(
\begin{matrix}
  \M A^{(11)} & 0 & 0 & 0
\\
  \M A^{(21)} & \M A^{(22)} & 0 & 0
\\
  \vdots & \cdots & \ddots & 0
\\
  \M A^{(m1)} & \M A^{(m2)} & \cdots & \M A^{(mm)}
\end{matrix}
\right),
\end{equation}
where $\M A^{(ab)}(x,\eps)$ is a sub-matrix of size $n_a \times n_b$.

We start by reducing {\bf diagonal blocks} of this matrix, $\M A^{(ab)}(x,\eps)$, to epsilon form.
This can be done by treating each diagonal block as an independent matrix, and proceeding as described in the previous sections.
Since the characteristic polynomial of a block-triangular matrix is a product of characteristic polynomials of its diagonal blocks, i.e.,
\begin{equation}
  \det(\M A(x,\eps)-\lambda \M I) =
    \det(\M A^{(11)}(x,\eps) - \lambda\M I) \cdot \ldots \cdot
    \det(\M A^{(mm)}(x,\eps) - \lambda \M I),
\end{equation}
once we have normalized the diagonal blocks, the whole matrix becomes normalized as well (but not necessarily Fuchsian yet).

Next, we fuchsify {\bf off-diagonal blocks} given by rectangular matrices $\M A^{(ab)}(x,\eps)$, $a > b$.
To do this, let us look at the parts of \eqref{eq:dj} related to such a block:
\begin{equation}
  \begin{cases}
  \partial_x \V J^{(a)} =
    \M A^{(aa)}\,\V J^{(a)} + \ldots \\
  \partial_x \V J^{(b)} =
    \M A^{(b\,a)}\,\V J^{(a)} +
    \M A^{(bb)}\,\V J^{(b)} + \ldots
  \end{cases}
\end{equation}
If $\M A^{(b\,a)}(x,\eps)$ has a singularity of Poincar\'e rank $r>0$ at $x=x_k$, then to reduce $r$ we can apply this basis transformation:
\begin{equation}
  \label{eq:offdiagt}
  \V J^{(b)} = \V J'^{(b)} + \Delta^{-r} \M D\, \V J^{(a)},
\end{equation}
where $\M D$ is some constant matrix, and $\Delta=x-x_k$ if $x_k\ne\infty$, or $\Delta=1/x$ if $x_k=\infty$.
This transformation changes $\M A^{(b\,a)}_{k0}(\eps)$ into
\begin{equation}
  \M A'^{(ba)}_{k0}(\eps) = \M A^{(ba)}_{k0}\!(\eps) + r\,\M D + \M A^{(bb)}_{k0}\!(\eps) \, \M D - \M D\, \M A^{(aa)}_{k0}(\eps)
\end{equation}

If both $\M A^{(aa)}_{k0}(\eps)$ and $\M A^{(bb)}_{k0}(\eps)$ have been factorized then it is always possible to solve the right-hand side of this equation for $\M D$, thus reducing the Poincar\'e rank of $\M A'^{(ba)}$ by one.
Moreover, this transformation only affects $\M A^{(bi)}$ for $i \le a$ and $\M A^{(ia)}$ for $i > b$, therefore if we will sequentially apply the transformation \eqref{eq:offdiagt} for each off-diagonal block, starting from the first row to the last, and from the last column to the first, we will obtain a fully Fuchsian (and still normalized) matrix.

Finally, we factorize the whole matrix as described in Section~\ref{sec:fact}, thus completing the transformation to epsilon form.

In \fuchsia the process of shuffling a matrix to its shortest lower block-diagonal form is performed by the \code{block\_triangular\_form} routine; fuchsification and normalization of diagonal blocks is done by the \code{reduce\_diagonal\_blocks}; fuchsification of the remaining off-diagonal blocks is done by the \code{fuchsify\_off\_diagonal\_blocks}.

\section{Using \fuchsia}
\label{sec:3}

\subsection{Installation}

To run \fuchsia one needs \sage~\cite{sagemath} version 7.0 or higher to be installed on the computer.
This task can be accomplished by following installation instructions available at the website \url{http://www.sagemath.org}.
\footnote{
    Some \linux distributions have \sage available in their package repositories; we do not recommend using those.
    A number of \maxima releases contain bugs which \fuchsia is sensitive to, and so far the official \sage builds have avoided those releases (unlike some \linux distributions).
}

\sage is a free and open-source Computer Algebra System licensed under GPL.
It is written in \python~2.7 and combines together a number of existing open-source mathematical systems and libraries like \texttt{Maxima}, \texttt{Singular}, and others with the goal of providing the best free CAS.
In particular our code heavily relies on the interface to \maxima~\cite{maxima}.

We also should obtain a file \code{fuchsia.py} which can be downloaded from the website \url{https://github.com/gituliar/fuchsia}, where many examples of usage and reduced matrices are also collected.

\vspace{10mm}

\subsection{Usage from the command line}

To run \fuchsia use the command\footnote{For brevity, we use a shortcut \code{fuchsia} which is equivalent to \code{sage -python fuchsia.py}.}
\begin{Verbatim}
    \prompt{$}{sage -python fuchsia.py <action> <options>}
\end{Verbatim}
where
\begin{itemize}
  \item \code{<action>} is one of the algorithms described in the previous section, i.e., \code{fuchsify}, \code{normalize}, \code{factorize}, or auxiliary action \code{transform}, which applies a user-defined transformation to the given matrix.
  \item \code{<options>} are action-dependent options described in the help message printed with the help of \code{fuchsia -h} command.
\end{itemize}

In the following we provide a complete help information printed by \code{fuchsia -h}:
\begin{Verbatim}
Fuchsia v16.11.14
Authors: Oleksandr Gituliar, Vitaly Magerya

Usage:
    fuchsia [-hv] [--use-maple] [-f <fmt>] [-l <path>] [-P <path>]
            <command> <args>...

Commands:
    reduce [-x <name>] [-e <name>] [-m <path>] [-t <path>] <matrix>
        find an epsilon form of the given matrix

    fuchsify [-x <name>] [-m <path>] [-t <path>] <matrix>
        find a transformation that will transform a given matrix
        into Fuchsian form

    normalize [-x <name>] [-e <name>] [-m <path>] [-t <path>] <matrix>
        find a transformation that will transform a given Fuchsian
        matrix into normalized form

    factorize [-x <name>] [-e <name>] [-m <path>] [-t <path>] <matrix>
        find a transformation that will make a given normalized
        matrix proportional to the infinitesimal parameter

    sort [-m <path>] [-t <path>] <matrix>
        find a block-triangular form of the given matrix

    transform [-x <name>] [-m <path>] <matrix> <transform>
        transform a given matrix using a given transformation

    changevar [-x <name>] [-m <path>] <matrix> <expr>
        transform matrix by susbtituting free variable by a
        given expression

Options:
    -h          show this help message
    -f <fmt>    matrix file format: mtx or m (default: mtx)
    -l <path>   write log to this file
    -v          produce a more verbose log
    -P <path>   save profile report into this file
    -x <name>   use this name for the free variable (default: x)
    -e <name>   use this name for the infinitesimal parameter (default: eps)
    -m <path>   save the resulting matrix into this file
    -t <path>   save the resulting transformation into this file
    --use-maple speed up calculations by using Maple when possible

Arguments:
    <matrix>    read the input matrix from this file
    <transform> read the transformation matrix from this file
    <expr>      arbitrary expression
\end{Verbatim}

Simple and more advanced results obtained with the help of \fuchsia are located in the directory \code{examples}.
There you will find many examples of original matrices together with generated transformations which lead to the epsilon form of corresponding matrices.
Another example of applying \fuchsia to find master integrals for next-to-leading order contributions to splitting functions in QCD were discussed in \cite{GM16}.

\subsection{Usage from \sage or \python}
\label{sec:usage_py}

You can also use \fuchsia as a library by starting the \sage prompt and importing the \texttt{fuchsia} module like this:

\begin{Verbatim}
    \prompt{$}{sage}
    ┌────────────────────────────────────────────────────────────────────┐
    │ SageMath Version 7.1, Release Date: 2016-03-20                     │
    │ Type "notebook()" for the browser-based notebook interface.        │
    │ Type "help()" for help.                                            │
    └────────────────────────────────────────────────────────────────────┘
    \prompt{sage:}{from fuchsia import *}
\end{Verbatim}

In order to give an example for the API, let us try to reduce a simple matrix.
For the list of functions available after import, please, read the next section.

\begin{Verbatim}
    \prompt{sage:}{x, eps = var("x eps")}
    \prompt{sage:}{M = matrix([}
    \prompt{....:}{  [(2-eps)/x, 0, 0],}
    \prompt{....:}{  [x/(x-1), eps/x, 0],}
    \prompt{....:}{  [(1+2*eps)/x**3, 0, (1+eps)/x/(x+1)]}
    \prompt{....:}{])}
\end{Verbatim}

First, let us see where the singularities of this matrix are located:

\begin{Verbatim}[commandchars=\\!|]
    \prompt!sage:|!singularities(M, x)|
    {-1: 0, 0: 2, 1: 0, +Infinity: 1}
\end{Verbatim}

So, 4 singularities in total, with the Poincar\'e rank being 2 at $\code{x}=0$, 1 at $\code{x}=\infty$ and 0 everywhere else.
To get rid of non-zero ranks (thus transforming the system into Fuchsian form) we will need to \textit{fuchsify} this matrix as:

\begin{Verbatim}[commandchars=\\!|]
    \prompt!sage:|!Mf, Tf = fuchsify(M, x)|
    \prompt!sage:|!Mf|
    [ -(eps - 2)/x           0                          0]
    [   -1/(x - 1)  (eps -1)/x                          0]
    [(2*eps + 1)/x           0  (eps + 2*x + 3)/(x^2 + x)]
    \prompt!sage:|!singularities(Mf, x)|
    {-1: 0, 0: 0, 1: 0, +Infinity: 0}
\end{Verbatim}

Now, let us take a look at the eigenvalues of \code{Mf} residues:

\begin{Verbatim}
    \prompt{sage:}{[matrix_residue(Mf, x, x0).eigenvalues()}
    \prompt{....:}{  for x0 in [-1, 0, 1, Infinity]]}
    [[-eps - 1, 0, 0],
     [-eps + 2, eps - 1, eps + 3],
     [0, 0, 0],
     [-eps + 1, eps - 2, -2]]
\end{Verbatim}

Many of these eigenvalues are not equal to zero in the limit $\code{eps}\to0$, so \code{Mf} is not normalized.
It is, however, the case that all of the eigenvalues are of the form $n + m*\code{eps}$, so there is a chance that we will be able to normalize \code{Mf}.
Let us try:

\begin{Verbatim}
    \prompt{sage:}{Mn, Tn = normalize(Mf, x, eps)}
    \prompt{sage:}{Mn}
    [-eps/x                                                            ...
    [(4*eps^3 - 8*eps^2 - (4*eps^2 - 6*eps + 3)*x + 5*eps)/((4*eps^3 - ...
    [((2*eps + 1)*x + 3*eps + 1)/(x^2 + x)                             ...
    \prompt{sage:}{[matrix_residue(Mn, x, x0).eigenvalues()}
    \prompt{....:}{  for x0 in [-1, 0, 1, Infinity]]}
    [[-eps, 0, 0],
     [-eps, eps, eps],
     [0, 0, 0],
     [-eps, eps, 0]]
\end{Verbatim}

So, the matrix is normalized, but it grew quite a bit larger.
This happens.
Sometimes it is possible to simplify it a bit:

\begin{Verbatim}
    \prompt{sage:}{Ms, Ts = simplify_by_jordanification(Mn, x)}
    \prompt{sage:}{Ms}
    [               -eps/x      0              0]
    [            1/(x - 1)  eps/x              0]
    [1/2*(eps + 1)/(x + 1)      0  eps/(x^2 + x)]
\end{Verbatim}

That is much better.

Finally, we need to \textit{factorize} \code{Ms} to complete the reduction:

\begin{Verbatim}
    \prompt{sage:}{Mr, Tr = factorize(Ms, x, eps)}
    \prompt{sage:}{Mr}
    [         -eps/x      0              0]
    [1/4*eps/(x - 1)  eps/x              0]
    [5/8*eps/(x + 1)      0  eps/(x^2 + x)]
\end{Verbatim}

This is the fully transformed matrix.
As you can see, it is both proportional to \code{eps} and Fuchsian.
To make sure we got everything right, we can double-check the full transformation:

\begin{Verbatim}
    \prompt{sage:}{T = (Tf*Tn*Ts*Tr).simplify_rational()}
    \prompt{sage:}{(Mr - transform(M, x, T)).is_zero()}
    True
\end{Verbatim}

Note that we have used the construct \code{(A - B).is\_zero()} to compare matrices instead of the more obvious \code{bool(A == B)}.
This is a \sage idiosyncrasy; the more obvious way compares symbolic matrices only structurally.

Of course, you do not need to walk through all these steps yourself every time.
Normally, you just need to call this one function to do all the reduction work:

\begin{Verbatim}
    \prompt{sage:}{MM, TT = epsilon_form(M, x, eps)}
    \prompt{sage:}{MM}
    [                          -eps/x      0              0]
    [                 1/4*eps/(x - 1)  eps/x              0]
    [1/4*(9*eps*x + 13*eps)/(x^2 + x)      0  eps/(x^2 + x)]
    \prompt{sage:}{(MM - transform(M, x, TT)).is_zero()}
    True
\end{Verbatim}

Notice that this matrix is slightly more complex than the one we have obtained step by step above.
This also happens.
The final form we are computing is not unique, and it will be different depending on the precise sequence of reduction steps you have taken.

Additionally, many of the transformations take a special $seed$ parameter to control the order of operations they perform internally.
By supplying different seeds, you will obtain different results as well.

\subsubsection{Function reference}

\begin{description}[style=nextline]

\functionitem{epsilon\_form}{\M M, x, epsilon, seed=0}
Fully reduces a system of equations defined by a matrix $\M M$, an independent variable $x$, and an infinitesimal parameter $epsilon$.
Returns a pair of values: the transformed matrix $\M M'$ and the transformation matrix $\M T$.
Raises $FuchsiaError$, if the system is irreducible.

The reduction is performed by first converting $\M M$ to block-triangular form, then reducing the diagonal blocks via $\F{fuchsify}$, $\F{normalize}$ and $\F{factorize}$, reducing off-diagonal blocks as described in Section~\ref{sec:blockreduce}, and finally factorizing $epsilon$ via $\F{factorize}$.

\functionitem{fuchsify}{\M M, x, seed=0}
Reduces a system defined by a matrix $\M M$ and an independent variable $x$ to Fuchsian form.
That is, it makes sure that the Poincar\'e ranks of all singularities of the transformed matrix $\M M'$ are $0$.
Returns a pair of values: the transformed matrix $\M M'$ and the transformation $\M T$.
If the system is irreducible, it raises $FuchsiaError$.

\functionitem{normalize}{\M M, x, epsilon, seed=0}
Transforms a Fuchsian system defined by a matrix $\M M$, an independent variable $x$ and, an infinitesimal parameter $epsilon$ to a normalized form.
That is, it makes sure that real parts of the eigenvalues of all matrix residues of the transformed matrix $\M M'$ lie in the range $[-1/2, 1/2)$ in the limit $epsilon\to0$.
Returns a pair of values: the transformed matrix $\M M'$ and the transformation $\M T$.
If such a transformation can not be found, it raises $FuchsiaError$.

\functionitem{factorize}{\M M, x, epsilon, seed=0}
Transforms a normalized system defined by a matrix $\M M$, an independent variable $x$, and an infinitesimal parameter $epsilon$ so that the
transformed matrix $\M M'$ is proportional to $epsilon$.
Returns a pair of values: the transformed matrix $\M M'$ and the transformation $\M T$.
If such a transformation can not be found, it raises $FuchsiaError$.

\functionitem{block\_triangular\_form}{\M M}
Transforms a matrix $\M M$ into a lower block-triangular form.

Returns three values: a transformed matrix $\M M'$, a transformation matrix $\M T$, and a list of tuples $(m_i, n_i)$, where $n_i$ is the size of $i$-th diagonal block, and $m_i = \sum_{j=1}^{i-1} n_j$.
The tuple list represents block structure of $\M M'$; it is used by the next two functions, where it is passed as the argument $B$.

\functionitem{reduce\_diagonal\_blocks}{\M M, x, epsilon, B=None, seed=0}
Finds a transformation that reduces diagonal blocks of $\M M$ into epsilon form.
If $B$ is not provided, it transforms $\M M$ into lower triangual form before reduction.
Otherwise it assumes $\M M$ blocks are described by $B$.

Returns two values: a transformed matrix $\M M$ and a transformation matrix $\M T$.

\functionitem{fuchsify\_off\_diagonal\_blocks}{\M M, x, epsilon, r=None}
Given a matrix $M$ with diagonal blocks in epsilon form, it transforms off-diagonal blocks in to Fuchsian form.
If $B$ is not provided, it transforms $\M M$ into lower triangual form before reduction.
Otherwise it assumes that blocks of $\M M$ are described by $B$.

Returns two values: a transformed matrix $\M M$ and a transformation matrix $\M T$.

\functionitem{simplify\_by\_jordanification}{\M M, x}
Tries to simplify a system defined by a matrix $\M M$ and an independent variable $x$ by constant transformations that transform leading expansion coefficients of $\M M$ into their Jordan forms.
Returns a pair of values: the simplified matrix $\M M'$ and the transformation $\M T$.
If none of the attempted transformations reduces the complexity of $\M M$ (as measured by $\F{matrix\_complexity}$), it returns the original matrix and the identity transformation.

\functionitem{simplify\_by\_factorization}{\M M, x}
Tries to simplify a system defined by a matrix $\M M$ and an independent variable $x$ by a constant transformation that extracts common factors found in $\M M$ (if any).
Returns a pair of values: the simplified matrix $\M M'$ and the transformation $\M T$.

\functionitem{matrix\_complexity}{\M M}
This function is used as a measure of matrix complexity by $\F{fuchsify}$ and simplification functions.
Currently it is defined as the length of textual representation of matrix $\M M$.

\functionitem{balance}{\M P, x_1, x_2, x}
Returns a \textit{balance} transformation between points $x=x_1$ and $x=x_2$ using the projector matrix $\M P$.
See eq.~\eqref{eq:bal} for the definition of a \textit{balance}.

\functionitem{transform}{\M M, x, \M T}
Transforms a system defined by a matrix $\M M$ and an independent variable $x$ using the transformation matrix $\M T$ as specified by eq.~\eqref{eq:ta}.
Returns the transformed matrix $\M M'$.

\functionitem{balance\_transform}{\M M, \M P, x_1, x_2, x}
Same as $\F{transform}(\M M, x, \F{balance}(\M P, x_1, x_2, x))$, but implemented more efficiently: since the inverse of $\F{balance}(\M P, x_1, x_2, x)$ is $\F{balance}(\M P, x_2, x_1, x)$, this function can avoid a time-consuming matrix inversion operation that $\F{transform}$ must perform.

\functionitem{singularities}{\M M, x}
Finds values of $x$ around which the matrix $\M M$ has a singularity in $x$.
Returns a dictionary with $\{x_i: p_i\}$ entries, where $p_i$ is the Poincar\'e rank of $\M M$ at $x=x_i$.
The set of singular points can include \code{Infinity}, if $\M M$ has a singularity at $x\to\infty$.

\functionitem{matrix\_c0}{\M M, x, x_0, p}
Returns the 0-th coefficient of the series expansion of a matrix $\M M$ around $x=x_0$, assuming the Poincar\'e rank of $\M M$ at that point is $p$.
If $x_0$ is \code{Infinity}, it returns minus the coefficient at the highest power of $x$.
In other words, it returns $\M A_{k0}$ from eq.~\eqref{eq:axeps}.

\functionitem{matrix\_c1}{\M M, x, x_0, p}
Returns the 1-th coefficient of the series expansion of a matrix $\M M$ around $x=x_0$, assuming the Poincar\'e rank of $\M M$ at that point is $p$.
If $x_0$ is \code{Infinity}, it returns minus the coefficient at the second-to-highest power of $x$.
In other words, it returns $\M A_{k1}$ from eq.~\eqref{eq:axeps}.

\functionitem{matrix\_residue}{\M M, x, x_0}
Returns a residue of a matrix $\M M$ at $x=x_0$, assuming that the Poincar\'e rank of $\M M$ at $x=x_0$ is $0$.
Returns matrix residue at infinity if $x_0$ is \code{Infinity}.

This is the same as $\F{matrix\_c0}(\M M, x, x_0, 0)$.

\functionitem{export\_matrix\_to\_file}{filename, \M M, fmt=\code{"mtx"}}
Writes a matrix $\M M$ to a file $filename$ using MatrixMarket array format if $fmt$ is \code{"mtx"} (which is the default), or Mathematica format if $fmt$ is \code{"m"}.

\functionitem{import\_matrix\_from\_file}{filename}
Reads a symbolic matrix from a file $filename$.
Both Mathematica and MatrixMarket array formats are supported.
The exact format will be autodetected.

\functionitem{setup\_fuchsia}{verbosity=0, use\_maple=\code{False}}
Modifies some of the \fuchsia inner workings.
In particular, it sets $verbosity$ to $2$ to enable verbose logging, $1$ to enable normal logging, and $0$ to only log errors.

Set $use\_maple$ to \code{True} to enable usage of \texttt{Maple} to speed up calculations when possible; this may be particularly beneficial for big matrices, and for matrices with singularities at complex points.

\classitem{FuchsiaError}
This is the class of exceptions raised by the \fuchsia routines.
It indicates the inability to perform the requested reduction. 

\end{description}

\section{Summary}
\label{sec:4}

In this paper we have presented \fuchsia, a program for reducing differential equations for Feynman master integrals to the epsilon form based on the Lee algorithm \cite{Lee15} which consists of three main computational steps: fuchsification, normalization, and factorization.
\fuchsia is open-source nature and depends on free software tools only: the programming language \python and the computer algebra system \maximasage, which makes it available for everyone.
Unfortunately, for some heavy-duty computations one needs to switch from \maxima to \maple, which is more powerful for some tasks, hence the access to latter is desirable.
Though it helps a lot in some situation, \maple is not a universal solution and is not able to process huge intermediate expressions which arise during the reduction process of some advanced examples.
It happens due to inability to work with polynomials which contain algebraic numbers in their coefficients (even in the case of complex coefficients the performance drastically decreases).

Despite of the discussed limitation \fuchsia shows a great performance in many cases.
It is possible due to the optimization for block-triangular (or sparse) matrices, which allows to reduce relatively large matrices: the reduction of ${74\times74}$ matrix with 20 rational and complex singular points and at most $3\times3$ coupled blocks takes about an hour on a laptop with Intel i5 CPU.

\section*{Acknowledgment}

We are gratefully thankful for advanced examples of differential equations provided by Roman Lee and Costas Papadopolous.
We also appreciate useful suggestions from Sven Moch during writing of this paper.

This work has been supported by the Deutsche Forschungsgemeinschaft in Sonderforschungs\-be\-reich 676 {\it Particles, Strings, and the Early Universe} and by the Narodowe Centrum Nauki with the Sonata Bis grant DEC-2013/10/E/ST2/00656.

\bibliography{fuchsia}

\end{document}